\documentclass[10pt,conference]{IEEEtran}
\usepackage{array}
\usepackage{cite}
\usepackage{calc}
\usepackage{caption}
\usepackage{graphicx}
\usepackage{psfrag}
\usepackage[tight,footnotesize]{subfigure}
\usepackage{stfloats}
\usepackage{url}
\usepackage{subfig}
\usepackage{longtable}
\usepackage{stfloats}
\usepackage{amsfonts,amssymb,amsbsy,bm,paralist,theorem,cite,ifthen,color}
\usepackage[cmex10]{amsmath}
\usepackage{algorithmic,algorithm}

\usepackage{multirow}

\newtheorem{Lemma}{Lemma}
\newtheorem{Proposition}{Proposition}

\newcommand\ab{\ensuremath{{\bm a}}}

\newcommand\eb{\ensuremath{{\bm e}}}
\newcommand\hb{\ensuremath{{\bm h}}}

\newcommand\vb{\ensuremath{{\bm v}}}
\newcommand\wb{\ensuremath{{\bm w}}}
\newcommand\xb{\ensuremath{{\bm x}}}

\newcommand\Ib{\ensuremath{{\bm I}}}
\newcommand\Qb{\ensuremath{{\bm Q}}}

\newcommand\Yb{\ensuremath{{\bm Y}}}

\newcommand\tr{\ensuremath{{\rm Tr}}}

\newcommand\zerob{\ensuremath{{{\bf 0}}}}
\newcommand\Phib{\ensuremath{{{\bm \Phi}}}}

\begin{document}
\bibliographystyle{IEEEtran}

\title{A {Low-Complexity} Algorithm for Worst-Case Utility Maximization in Multiuser MISO {Downlink}}


\author{Kun-Yu Wang$^\star$, Haining Wang$^\dag$, Zhi Ding$^\dag$, and Chong-Yung
Chi$^\star$\\~\\
\begin{tabular}{cc}
$^\star$Institute of Communications Engineering       & $^\dag$Department of Electrical and Computer Engineering\\
National Tsing Hua University,                        & University of California, Davis,\\
Hsinchu, Taiwan 30013                                 & Davis, CA 95616       \\
\small E-mail: kunyuwang7@gmail.com,~cychi@ee.nthu.edu.tw   & \small
E-mail: \{whnzinc,~zding.ucdavis\}@gmail.com
\end{tabular}
\vspace{-0.07cm}}

\maketitle
\begin{abstract}
This work considers {worst-case utility maximization (WCUM)} problem
for a downlink wireless system where a multi-antenna base station
communicates with multiple single-antenna users. Specifically, we
jointly design transmit covariance matrices for each user to
robustly maximize the worst-case (i.e., minimum) system utility
{function} under channel estimation errors bounded within a
spherical region. {This problem has been shown to be NP-hard, and so
any algorithms for finding the optimal solution may suffer from
prohibitively high complexity.} In view of this, we {seek} an
efficient and {more accurate} suboptimal solution for the {WCUM
problem}. {A low-complexity iterative WCUM algorithm is proposed for
this nonconvex problem by solving two convex problems
alternatively.} We also show the convergence of {the} proposed
algorithm, and prove its Pareto optimality {to} the {WCUM} problem.
{Some simulation results are presented to demonstrate its
substantial performance gain and higher computational efficiency
over existing algorithms.}
\end{abstract}
\IEEEpeerreviewmaketitle

\section{Introduction}

Linear transmit precoding has been recognized as an important
technique for capacity improvement {and} low implementation
complexity. Considering a transmit design for system utility
maximization in a single-cell multiuser multiple-input single-output
(MISO) wireless system, several works have focused on finding the
optimal beamforming solution for the problem
\cite{Liu2012,Joshi2012,Bjornson2012}. However, since the problem is
NP-hard in general \cite{Luo2008}, the computation complexity {of
any algorithms} for finding the optimal solution can be
prohibitively high, {rendering} the convergence rate very slow.
Therefore, those algorithms may be infeasible for real-time
implementation. In light of this, considerable attention has been
shifted towards finding a {more accurate and computationally
efficient solution} for the system utility maximization problem.
Assuming that the base station (BS) can perfectly acquire the
channel state information (CSI) of the users, some efficient
suboptimal algorithms have been proposed for the utility
maximization problem \cite{Ng2010,Shi2011,Tran2012}. The authors in
\cite{Tran2012} propose a suboptimal algorithm based on the idea of
successive convex approximation (SCA) for a weighted sum rate
maximization problem, with numerical results showing that {their}
algorithm outperforms existing ones.

However, in practical situations, it is inevitable to have channel
estimation errors at the BS due to finite {training resource (e.g.,
power and signal length)} or limited feedback bandwidth
\cite{Love2008}. If one uses inaccurate CSI directly to design the
transmit precoders, then the system performance may be degraded
{seriously}. In view of this, we design a transmit precoder by
taking CSI errors into account. Also, in a slow fading channel,
{under a strict} constraint on quality of service (QoS), the system
must be designed for the worst-case scenario \cite{Wiesel2007}. In
this paper, the CSI errors are assumed be bounded spherically, and
the transmit covariance matrices for {all the users} are designed to
maximize the worst-case (i.e., minimum) system utility function
against any possible CSI error, subject to a total transmit power
constraint. However, the worst-case utility maximization {(WCUM)}
problem is nonconvex and is generally {hard} to efficiently solve.
{Therefore}, very few efficient algorithms have been {reported} for
the {WCUM} problem.

One approach {to} developing a suboptimal solution for the {WCUM}
problem is based on SCA. Although the SCA-based algorithm presented
in \cite{Tran2012} relies on the assumption that the BS has perfect
CSI of the users, the SCA-based concept can be easily extended for
handling the {WCUM} problem. Specifically, we can apply a
{conservative convex approximation} to the {WCUM} problem, and then
iteratively update the associated {parameters} to improve the
performance. Since a conservative approximation is used in SCA-based
algorithms, performance of such algorithm would be degraded. {Hence
in this paper}, we propose a {low-complexity} algorithm for the
{WCUM} problem {without any conservative approximations}. In the
proposed {WCUM} algorithm, two convex optimization problems are
solved alternatively. The proposed algorithm is guaranteed to
converge, and the limit point is Pareto optimal to the {WCUM}
problem. {Some simulation results are presented to show that the
proposed WCUM algorithm significantly outperforms the SCA-based
algorithm, and the former has lower computational complexity than
the later.}

\section{System Model and Problem Statement}\label{Sec: system model}

\subsection{System Model}

We consider a single-cell multiuser multiple-input single-output
(MISO) wireless system, where a BS equipped with $N_t$ antennas
communicates with $K$ single-antenna users. The transmitted signal
at the BS is {given by}
\begin{align}\label{TX signal}
\xb(t)=\sum_{k=1}^{K}\xb_{k}(t),
\end{align}
where $\xb_{k}(t)\in\mathbb{C}^{N_{t}}$ denotes the
information-bearing signal for {the} $k$th user. With \eqref{TX
signal}, the received signal at {the} $k$th user can be represented
as
\begin{align}
y_{k}(t)=\hb_{k}^{H}\xb(t)+n_{k}(t),
\end{align}
where $\hb_{k}\in\mathbb{C}^{N_t}$ denotes the channel vector
between the BS and {the} $k$th user, and $n_{k}(t)$ is the additive
noise at {the} $k$th user with power $\sigma_{k}^{2}>0$.

Assuming that $\xb_{k}(t)$ is complex Gaussian distributed with zero
mean and covariance matrix $\Qb_{k}\succeq\zerob$ (positive
semidefinite (PSD) matrix), i.e.,
$\xb_{k}(t)\sim\mathcal{CN}(\zerob,\Qb_{k})$, and considering
single-user detection, the achievable rate of {the} $k$th user can
be represented as (in bits/sec/Hz):
\begin{align}\label{capacity}
\!\!\!R_{k}(\{\Qb_{i}\}_{i=1}^{K},\hb_{k})=\log_{2}\bigg(1+\frac{\hb_{k}^{H}\Qb_{k}\hb_{k}}{\sum_{\ell\neq
k}^{K}\hb_{k}^{H}\Qb_{\ell}\hb_{k}+\sigma_{k}^{2}}\bigg).
\end{align}
The goal of the transmit precoding is to design the transmit
covariance matrices $\{\Qb_{k}\}_{k=1}^{K}$ such that {a} system
utility function $U(R_{1},\ldots,R_{K})$ can be maximized, {as
formulated as the following optimization problem:}
\begin{subequations}\label{Conventional problem}
\begin{align}
\max_{\substack{\Qb_{k}\in\mathbb{H}^{N_{t}},\\k=1,\ldots,K}}&~
U(\{R_{k}(\{\Qb_{i}\}_{i=1}^{K},\hb_{k})\}_{k=1}^{K})\\
{\rm s.t.}&~\sum_{k=1}^{K} \tr(\Qb_{k})\leq
P,~\Qb_{1},\ldots,\Qb_{K}\succeq\zerob,
\end{align}
\end{subequations}
where {$\tr(\cdot)$ denotes the trace of a matrix}, and $P>0$ is a
preset maximum allowed total transmit power. We assume the system
utility function $U(R_{1},\ldots,R_{K})$ to be strictly increasing
and concave {with respect to $R_{k}$, for $k=1,\ldots,K$}, {as
satisfied by many practical system performance measures}
\cite{Luo2008}, e.g., sum-rate utility
$U(R_{1},\ldots,R_{K})=(1/K)\sum_{k=1}^{K}R_{k}$. Although problem
\eqref{Conventional problem} is not convex in general {due to the
nonconcave utility function with respect to
$\{\Qb_{k}\}_{k=1}^{K}$}, it {falls in} the class of monotonic
optimization problems \cite{Rubinov2001}, and the optimal transmit
covariance matrices can be obtained
\cite{Liu2012,Joshi2012,Bjornson2012}, {in spite of high
computational complexity}.

\subsection{Problem Statement}

In practice, the BS cannot perfectly acquire the CSI from users due
to finite training power or limited feedback bandwidth
\cite{Love2008}. In this work, the true channel is modeled {as}
\begin{align}\label{channel model}
\hb_{k}=\hat{\hb}_{k}+\eb_{k},
\end{align}
where $\hat{\hb}_{k}\in\mathbb{C}^{N_t}$ denotes the channel
estimate of $\hb_{k}$ at the BS, and $\eb_{k}\in\mathbb{C}^{N_t}$ is
the corresponding CSI error vector which is assumed to lie in a
{norm ball} with radius $r_{k}>0$, i.e., $\|\eb_{k}\|\leq r_{k}$,
where $||\cdot||$ denotes the vector 2-norm.

Our goal is to jointly design the transmit covariance matrices
$\{\Qb_{k}\}_{k=1}^{K}$ such that the worst-case (i.e., minimum)
system utility over the CSI errors is maximized, subject to a total
transmit power constraint. Mathematically, the {WCUM} problem can be
{formulated} as
\begin{subequations} \label{designed problem 2}
\begin{align}
\max_{\substack{\Qb_{k}\in\mathbb{H}^{N_{t}},\\k=1,\ldots,K}}
&~\min_{\substack{\|\eb_{k}\|\leq r_{k},\\k=1,\ldots,K}}
U\big(\{R_{k}(\{\Qb_{i}\}_{i=1}^{K},{\hat
\hb}_{k}+\eb_{k})\}_{k=1}^{K}\big) \\
{\rm s.t.}&~ \sum_{k=1}^{K} \tr(\Qb_{k})\leq
P,~\Qb_{1},\ldots,\Qb_{K}\succeq\zerob, \label{designed problem
constraint 21}
\end{align}
\end{subequations}
{Equivalently, problem \eqref{designed problem 2} can be
reformulated as}
\begin{center}
\fbox{\parbox[]{0.97 \linewidth}{\vspace{-0.3cm}
\begin{subequations}\label{designed problem 2 reformulation 2}
\begin{align}
\max_{\substack{\Qb_{k}\in\mathbb{H}^{N_{t}},t_{k}\in\mathbb{R},\\k=1,\ldots,K}}
&~ U(t_{1},\ldots,t_{K})\\
{\rm s.t.} &~
\frac{(\hat{\hb}_{k}+\eb_{k})^{H}\Qb_{k}(\hat{\hb}_{k}+\eb_{k})}{\sum_{\ell\neq
k}^{K}(\hat{\hb}_{k}+\eb_{k})^{H}\Qb_{\ell}(\hat{\hb}_{k}+\eb_{k})+\sigma_{k}^{2}}\!\geq\!
t_{k},\notag\\
&~\forall\|\eb_{k}\|\leq r_{k},~k=1,\ldots,K,\label{designed problem 2 reformulation 21}\\
&~ \sum_{k=1}^{K} \tr(\Qb_{k})\leq
P,~\Qb_{1},\ldots,\Qb_{K}\succeq\zerob, \label{designed problem 2
reformulation 22}
\end{align}
\end{subequations}
\vspace{-0.2cm}}}
\end{center}
{where we have made the change of variables
\begin{align}
R_{k}(\{\Qb_{i}\}_{i=1}^{K},{\hat
\hb}_{k}+\eb_{k})=\log_{2}(1+t_{k})
\end{align}
in the utility function.} Problem \eqref{designed problem 2}, or
equivalently problem \eqref{designed problem 2 reformulation 2},
provides a maximum lower bound on the system utility function over
the CSI errors, given that $\{\hat{\hb}_{k}\}_{k=1}^{K}$ is known at
the BS. However, problem \eqref{designed problem 2 reformulation 2}
is hard to solve due to the infinitely many nonconvex constraints in
\eqref{designed problem 2 reformulation 21}. {Before presenting the
algorithm for efficiently and effectively handling problem
\eqref{designed problem 2 reformulation 2}, let us present a
simulation example to demonstrate the essentiality of the WCUM
design.}

\begin{figure}[t]
\vspace{-0.3cm}
\begin{center}
    \resizebox{0.53\textwidth}{!}{
    \psfrag{r}[Bc][Bc]{\large $r$}\hspace{-0.8cm}
        \includegraphics{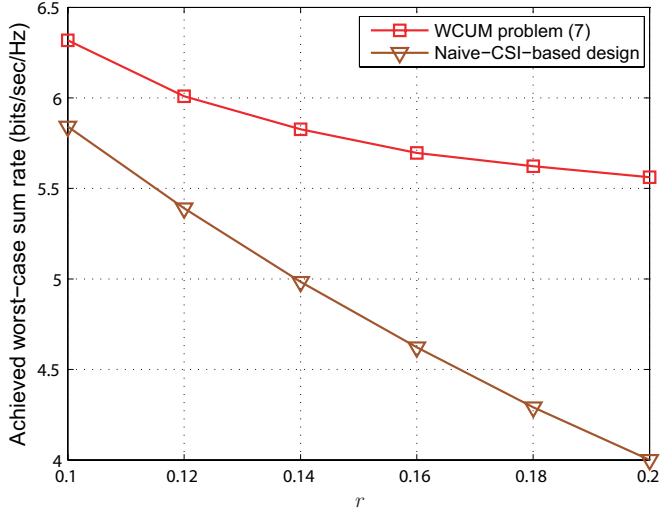}} 
        \caption{Achieved worst-case sum rate vs. CSI error radius $r$.
        }\label{Worst_Rate_Actual}\vspace{-0.45cm}
\end{center}
\end{figure}

\textbf{Example:} Consider sum-rate utility for problem
\eqref{designed problem 2 reformulation 2} and naive-CSI-based
design. For naive-CSI-based design, the transmit covariance matrices
are obtained by solving the conventional perfect-CSI-based problem
\eqref{Conventional problem}, where the channel estimates $\{{\hat
\hb}_{k}\}_{k=1}^{K}$ are used as if they were true channels. {For
each} realization of the channel estimates $\{{\hat
\hb}_{k}\}_{k=1}^{K}$ {generated with complex Gaussian
distribution}, the {optimal} transmit covariance matrices for
problem \eqref{designed problem 2 reformulation 2} and
naive-CSI-based designs {are obtained} by applying
branch-reduce-and-bound (BRB) algorithm {reported} in
\cite{Bjornson2012}. With the obtained transmit covariance matrices
and the presumed channels $\{{\hat \hb}_{k}\}_{k=1}^{K}$, the
worst-case sum rates were determined {to be the minimum value of
$R_{k}$ computed by \eqref{capacity} over $10^{6}$ true channels}
$\{\hb_{k}=\hat{\hb}_{k}+\eb_{k}\}_{k=1}^{K}$, where {the simulated
CSI errors $\{\eb_{k}\}_{k=1}^{K}$} are randomly and independently
generated satisfying $\|\eb_{k}\|\leq r$ for $k=1,\ldots,K$.
Figure~\ref{Worst_Rate_Actual} {displays some simulation results of}
the achieved worst-case sum rate versus $r$ {for $N_{t}=K=2$, $P=10$
dB, and $\sigma_{1}^{2}=\cdots=\sigma_{K}^{2}=0.01$, where each
result was} obtained by averaging over 100 realizations of the
channel estimates {$\{{\hat \hb}_{k}\}_{k=1}^{K}$}. From this
figure, one can see that improper transmit covariance matrices can
{yield} a very low sum rate in the worst case, especially for large
CSI error radius. \hfill $\blacksquare$

Since problem \eqref{designed problem 2 reformulation 2} is NP-hard
in general \cite{Luo2008}, {the computational complexity of any
algorithms for finding the optimal solution of problem
\eqref{designed problem 2 reformulation 2} can be prohibitively
high,} and thus it is infeasible for real-time implementation
\cite{Liu2012,Joshi2012,Bjornson2012}. {Next, we concentrate on the
proposed low-complexity algorithm for finding a more accurate
suboptimal solution for problem \eqref{designed problem 2
reformulation 2}.}

\section{A {Low-Complexity WCUM Algorithm}}\label{Sec: Fast
Algorithms}

{In this section, we first present the proposed iterative algorithm
for problem \eqref{designed problem 2 reformulation 2}, and then
prove its convergence.}

\subsection{Proposed Algorithm}

To proceed, let us apply S-lemma \cite{Boyd2004} to constraint
\eqref{designed problem 2 reformulation 21}, and then problem
\eqref{designed problem 2 reformulation 2} can be equivalently
reformulated as (see, e.g., \cite{Bjornson2012} for details):
\begin{subequations}\label{designed problem 2 reformulation N=1}
\begin{align}
\max_{\substack{\Qb_{k}\in\mathbb{H}^{N_{t}},\\t_{k},
\lambda_{k}\in\mathbb{R},\\k=1,\ldots,K}}&~U(t_{1},\ldots,t_{K})\\
{\rm s.t.}&~\Phib_{k}\big(t_{k}, \lambda_{k},
\{\Qb_{i}\}_{i=1}^{K}\big)
\succeq \zerob,~k=1,\ldots,K,\label{designed problem 2 reformulation N=1 1}\\
&~\sum_{k=1}^{K}\tr(\Qb_{k})\leq P, \label{designed problem 2 reformulation N=1 2}\\
&~\lambda_{k}\geq 0, ~\Qb_{k}\succeq\zerob, ~k=1,\ldots,K,
\end{align}
\end{subequations}
where $\lambda_{1},\ldots,\lambda_{K}\in\mathbb{R}$ are the
introduced slack variables, and
\begin{align}\label{Defn: Phi}
&\Phib_{k}\big(t_{k}, \lambda_{k},
\{\Qb_{i}\}_{i=1}^{K}\big)\triangleq
\begin{bmatrix} \Ib_{N_{t}} \\ {\hat \hb}_{k}^{H}
\end{bmatrix}
\bigg(\Qb_{k}-t_{k}\sum_{\ell\neq k}\Qb_{\ell}\bigg)
\begin{bmatrix}
\Ib_{N_{t}} \\ {\hat \hb}_{k}^{H}
\end{bmatrix}^{H}\notag\\
&\qquad~~~~~+
\begin{bmatrix}
\lambda_{k}\Ib_{N_{t}} & \zerob\\
\zerob & -\lambda_{k}r_{k}^{2}-t_{k}\sigma_{k}^{2}
\end{bmatrix},~k=1,\ldots,K.
\end{align}
Although problem \eqref{designed problem 2 reformulation N=1} is
still not convex due to nonconvex constraint \eqref{designed problem
2 reformulation N=1 1}, the problem has a more tractable form than
problem \eqref{designed problem 2 reformulation 2}. To {develop the}
proposed algorithm, we need the following lemma.

\begin{Lemma}\label{Lemma: eigenvalue decreasing}
Each eigenvalue of the matrix $\Phib_{k}(t_{k}, \lambda_{k},
\{\Qb_{i}\}_{i=1}^{K})$ {defined in \eqref{Defn: Phi}} decreases
with $t_{k}$, for $k=1,\ldots,K$.
\end{Lemma}

\emph{Proof:} By letting $\delta_{k,j}\in\mathbb{R}$,
$j=1,\ldots,N_{t}+1$, be the eigenvalues of the matrix
$\Phib_{k}(t_{k}, \lambda_{k}, \{\Qb_{i}\}_{i=1}^{K})$ and denoting
$\vb_{k,j}\in\mathbb{C}^{N_{t}+1}$ as the {associated unit-norm}
eigenvector, we have
\begin{align}
\delta_{k,j}=&~\vb_{k,j}^{H}\Phib_{k}\big(t_{k}, \lambda_{k},
\{\Qb_{i}\}_{i=1}^{K}\big)\vb_{k,j} \nonumber\\
=&~\vb_{k,j}^{H}\Yb_{k,k}\vb_{k,j}+\lambda_{k}\vb_{k,j}^{H}
\begin{bmatrix}
\Ib_{N_{t}} & \zerob\\
\zerob & -r_{k}^{2}
\end{bmatrix}\vb_{k,j}\notag\\
&~ -t_{k}\bigg(\sum_{\ell\neq k}\vb_{k,j}^{H}
\Yb_{k,\ell}\vb_{k,j}+\sigma_{k}^{2}|[\vb_{k,j}]_{N_{t}+1}|^{2}\bigg),\label{eigenvalues}
\end{align}
where $[\ab]_{j}$ denotes $j$th entry of a vector $\ab$, and
\begin{align}
\Yb_{k,\ell}\triangleq
\begin{bmatrix}
\Ib_{N_{t}} \\ {\hat \hb}_{k}^{H}
\end{bmatrix}
\Qb_{\ell}
\begin{bmatrix}
\Ib_{N_{t}} \\ {\hat \hb}_{k}^{H}
\end{bmatrix}^{H}\succeq\zerob,~\ell=1,\ldots,K.\label{Defn: Y}
\end{align}
Since $\Yb_{k,\ell}$ is a PSD matrix, from \eqref{eigenvalues}, one
can easily show that the eigenvalue $\delta_{k,j}$ decreases with
$t_{k}$, for $j=1,\ldots,N_{t}+1$. This proof is thus complete.
\hfill $\blacksquare$

Problem \eqref{designed problem 2 reformulation N=1} aims to
maximize $t_{k}$ ({because} the utility function is strictly
increasing in $t_{k}$). According to Lemma \ref{Lemma: eigenvalue
decreasing}, {one} can maximize $t_{k}$ by maximizing the minimum
eigenvalue of the matrix $\Phib_{k}(t_{k}, \lambda_{k},
\{\Qb_{i}\}_{i=1}^{K})$ in \eqref{designed problem 2 reformulation
N=1 1} (which only involves $t_{k}$). Also, {it can be inferred}
that the minimum eignevalue of {$\Phib_{k}(t_{k}, \lambda_{k},
\{\Qb_{i}\}_{i=1}^{K})$} must be zero as the optimal solution of
problem \eqref{designed problem 2 reformulation N=1} is achieved.
Based on these facts, {let us consider the following iterative
approach for dealing with problem \eqref{designed problem 2
reformulation N=1}. At the $m$th iteration, $t_{1},\ldots,t_{K}$ are
updated by solving}
\begin{subequations}\label{decomposed problem 2}
\begin{align}
\max_{\substack{t_{k}\in\mathbb{R},\\k=1,\ldots,K}}&~
U(t_{1},\ldots,t_{K})\\
{\rm s.t.} &~\Phib_{k}\big(t_{k}, \lambda_{k}^{(m)},
\{\Qb_{i}^{(m)}\}_{i=1}^{K}\big) \succeq \zerob,~\forall k.
\label{decomposed problem 21}
\end{align}
\end{subequations}
{Let us denote the obtained optimal utility value in problem
\eqref{decomposed problem 2} as} ${\tilde
U}^{(m)}(\{\Qb_{k}^{(m)},\lambda_{k}^{(m)}\}_{k=1}^{K})$, in which
$\Qb_{k}^{(m)}$ and $\lambda_{k}^{(m)}$, $k=1,\ldots,K$, are
obtained by solving
\begin{subequations}\label{decomposed problem 1}
\begin{align}
\!\!\max_{\substack{\Qb_{k}\in\mathbb{H}^{N_{t}},\\
\lambda_{k},z_{k}\in\mathbb{R},\\k=1,\ldots,K}}
&~ {\Psi\triangleq} \sum_{k=1}^{K}z_{k}\\
{\rm s.t.} &~\Phib_{k}\big(t_{k}^{(m\!-\!1)}, \lambda_{k},
\{\Qb_{i}\}_{i=1}^{K}\big)
\!-\!z_{k}\Ib_{N_{t}+1}\!\succeq\! \zerob,~\forall k, \label{decomposed problem 11}\\
&~\sum_{k=1}^{K}\tr(\Qb_{k})\leq P,\\
&~\lambda_{k}\geq 0 ~\Qb_{k}\succeq\zerob, ~z_{k}\geq0,
~k=1,\ldots,K.
\end{align}
\end{subequations}
{The} obtained {optimal value of $\Psi$} {in problem
\eqref{decomposed problem 1} is denoted as}
$\Psi^{(m)}(\{t_{k}^{(m-1)}\}_{k=1}^{K})$, in which $t_{k}^{(m-1)}$,
$k=1,\ldots,K$, represent the solution obtained by solving problem
\eqref{decomposed problem 2} from the $(m-1)$th iteration. Problems
\eqref{decomposed problem 2} and \eqref{decomposed problem 1} are
convex, and thus can be efficiently solved. Specifically, since the
design variables $t_{1},\ldots,t_{K}$ in the objective function and
in the constraints of problem \eqref{decomposed problem 2} are
decoupled, and the objective function is strictly increasing in
$t_{k}$, for $k=1,\ldots,K$, by Lemma \ref{Lemma: eigenvalue
decreasing}, the optimal $\{t_{k}\}_{k=1}^{K}$ can be separately
obtained by simple bisection search; i.e., find a $t_{k}$ such that
the minimum eigenvalue of $\Phib_{k}(t_{k}, \lambda_{k}^{(m)},
\{\Qb_{i}^{(m)}\}_{i=1}^{K})$ in \eqref{decomposed problem 21} {is
equal to zero. The obtained WCUM algorithm} for problem
\eqref{designed problem 2 reformulation 2}, or equivalently problem
\eqref{designed problem 2 reformulation N=1}, is summarized in
Algorithm \ref{Algo: proposed algorithm}.

{\renewcommand{\baselinestretch}{1}
\begin{algorithm}[t]
\caption{Proposed {WCUM} algorithm for problem \eqref{designed
problem 2 reformulation 2}.}
\begin{algorithmic}[1]\label{Algo: proposed algorithm}
\STATE {Obtain} a feasible point $\{t_{k}^{(0)}\}_{k=1}^{K}$
according to \eqref{Initial point alternation-based algo}; set a
solution accuracy $\epsilon>0$; and set iteration index $m=0$.
\REPEAT
\STATE {Update $m:= m+1$.}
\STATE {Obtain} $\{\Qb_{k}^{(m)},\lambda_{k}^{(m)}\}_{k=1}^{K}$ by
solving problem \eqref{decomposed problem 1}.
\STATE {Obtain} $\{t_{k}^{(m)}\}_{k=1}^{K}$ by solving problem
\eqref{decomposed problem 2}.
\UNTIL the predefined stopping criterion is met, e.g., $|{\tilde
U}^{(m)}-{\tilde U}^{(m-1)}|\leq \epsilon$.
\end{algorithmic}
\end{algorithm}}

An initial feasible point $\{t_{k}^{(0)}\}_{k=1}^{K}$ for solving
problem \eqref{decomposed problem 1} in the {first} iteration can be
obtained by finding a feasible point of {problem \eqref{designed
problem 2 reformulation 2}}. Let the transmit covariance matrices be
of rank one, i.e., $\Qb_{k}=\wb_{k}\wb_{k}^{H}$ for $k=1,\ldots,K$,
where $\wb_{k}\in\mathbb{C}^{N_{t}}$, $k=1,\ldots,K$, can be any
arbitrary vectors such that the power constraint in \eqref{designed
problem 2 reformulation 22} is satisfied, i.e.,
$\sum_{k=1}^{K}\|\wb_{k}\|^{2}\leq P$. Therefore, from
\eqref{designed problem 2 reformulation 21}, a feasible point
$\{t_{k}^{(0)}\}_{k=1}^{K}$ {is given by}
\begin{align}
\!\!\!t_{k}^{(0)}\!=&\frac{\min_{\|\eb_{k}\|\leq r_{k}}|({\hat
\hb}_{k}+\eb_{k})^{H}\wb_{k}|^{2}}{\max_{\|\eb_{k}\|\leq
r_{k}}\sum_{\ell\neq k}^{K}|({\hat
\hb}_{k}+\eb_{k})^{H}\wb_{\ell}|^{2}+\sigma_{k}^{2}}\notag\\
=&\frac{([|{\hat \hb}_{k}^{H}\wb_{k}|-r_{k}\|\wb_{k}\|]^{+})^{2}}
{\sum_{\ell\neq k}^{K}(|{\hat
\hb}_{k}^{H}\wb_{\ell}|+r_{k}\|\wb_{\ell}\|)^{2}+\sigma_{k}^{2}},~k=1,\ldots,K,\label{lower
bound t}
\end{align}
where $[a]^{+}\triangleq \max\{a,0\}$. {Letting}
\begin{align}
\wb_{k}=\sqrt{\frac{P}{K}}\frac{{\hat \hb}_{k}}{\|{\hat
\hb}_{k}\|},~k=1,\ldots,K,
\end{align}
{in \eqref{lower bound t} gives rise to}
\begin{align}
t_{k}^{(0)}=\frac{\frac{P}{K}\big([\|{\hat
\hb}_{k}^{H}\|-r_{k}]^{+}\big)^{2}} {\sum_{\ell\neq
k}^{K}\frac{P}{K}\left(\frac{|{\hat \hb}_{k}^{H}{\hat
\hb}_{\ell}|}{\|{\hat
\hb}_{\ell}\|}+r_{k}\right)^{2}+\sigma_{k}^{2}},~k=1,\ldots,K.\label{Initial
point alternation-based algo}
\end{align}

\subsection{Convergence of Algorithm \ref{Algo: proposed
algorithm}}

In the following, we will show the convergence of Algorithm
\ref{Algo: proposed algorithm} and the limit point {to be} Pareto
optimal to problem~\eqref{designed problem 2 reformulation 2}. To
this end, we need the following lemma:


\begin{Lemma} \label{Lemma: eigenvalue vs t}
If the objective value of problem \eqref{decomposed problem 1}
obtained in the $m$th iteration is positive, i.e., $\Psi^{(m)}>0$,
then the system utility value can be increased by solving problem
\eqref{decomposed problem 2}, i.e., ${\tilde U}^{(m)}> {\tilde
U}^{(m-1)}$, {and} the limit values
$z_{1}^{\star},\ldots,z_{K}^{\star}$ obtained by Algorithm
\ref{Algo: proposed algorithm} must be zero.
\end{Lemma}

\emph{Proof:} Since $\Psi^{(m)}=\sum_{k=1}^{K}z_{k}^{(m)}$ is
positive, {let us assume $z_{k}^{(m)}>0$ for some $k$} without loss
of generality. Therefore, from \eqref{decomposed problem 11}, we
have
\begin{align}
\Phib_{k}\big(t_{k}^{(m-1)}, \lambda_{k}^{(m)},
\{\Qb_{i}^{(m)}\}_{i=1}^{K}\big) \succ\zerob.
\end{align}
According to Lemma \ref{Lemma: eigenvalue decreasing}, we can always
find a value, say $t_{k}^{(m)}$, such that
$t_{k}^{(m)}>t_{k}^{(m-1)}$ {is} feasible to problem
\eqref{decomposed problem 2}. As a result, we have ${\tilde
U}^{(m)}> {\tilde U}^{(m-1)}$ since the objective function of
problem \eqref{decomposed problem 2} is strictly increasing in
$t_{k}$.

Next, {let us show that as Algorithm \ref{Algo: proposed algorithm}
converges,} the limit values $z_{1}^{\star},\ldots,z_{K}^{\star}$
are all zero by contradiction. If {$z_{k}^{\star}\neq 0$}, then we
can further increase the objective function of problem
\eqref{decomposed problem 2}, which contradicts the {premise} that
the algorithm has converged. This proof is thus complete. \hfill
$\blacksquare$

Since ${\tilde U}^{(m)}$ in Algorithm~\ref{Algo: proposed algorithm}
is monotonically increasing {in the iteration number $m$} [by Lemma
\ref{Lemma: eigenvalue vs t}] and {its} value is bounded above due
to finite total transmit power $P$, we can conclude that
Algorithm~\ref{Algo: proposed algorithm} {must converge}. In {the}
following proposition, {the Pareto optimality of Algorithm
\ref{Algo: proposed algorithm} to problem \eqref{designed problem 2
reformulation 2} is established.}

\begin{Proposition}\label{Prop: Pareto optimal}
The limit point
$\{\Qb_{k}^{\star},t_{k}^{\star},\lambda_{k}^{\star}\}_{k=1}^{K}$
{obtained} by Algorithm \ref{Algo: proposed algorithm} is Pareto
optimal to problem \eqref{designed problem 2 reformulation 2}.
\end{Proposition}

\emph{Proof:} To show that the limit point
$\{\Qb_{k}^{\star},t_{k}^{\star},\lambda_{k}^{\star}\}_{k=1}^{K}$ is
Pareto optimal to problem \eqref{designed problem 2 reformulation
2}, or equivalently problem \eqref{designed problem 2 reformulation
N=1}, we first show that the limit point is a feasible point of
problem \eqref{designed problem 2 reformulation N=1}. Since the
point $\{\Qb_{k}^{\star}, \lambda_{k}^{\star}, z_{k}^{\star},
t_{k}^{\star}\}_{k=1}^{K}$ is feasible to problems \eqref{decomposed
problem 2} and \eqref{decomposed problem 1}, we have
\begin{subequations}\label{equi. feasible set}
\begin{align}
&~\Phib_{k}\big(t_{k}^{\star}, \lambda_{k}^{\star},
\{\Qb_{i}^{\star}\}_{i=1}^{K}\big)
\succeq \zerob,~k=1,\ldots,K,\\
&~\sum_{k=1}^{K}\tr(\Qb_{k}^{\star})\leq P,\\
&~\lambda_{k}^{\star}\geq 0, ~\Qb_{k}^{\star}\succeq\zerob,
~k=1,\ldots,K,
\end{align}
\end{subequations}
where we use the fact $z_{1}^{\star}=\cdots = z_{K}^{\star} =0$ in
\eqref{equi. feasible set} according to Lemma \ref{Lemma: eigenvalue
vs t}. Comparing \eqref{equi. feasible set} with the feasible set of
problem \eqref{designed problem 2 reformulation N=1}, one can
conclude that the limit point $\{\Qb_{k}^{\star},
\lambda_{k}^{\star}, t_{k}^{\star}\}_{k=1}^{K}$ generated by
Algorithm \ref{Algo: proposed algorithm} is feasible to problem
\eqref{designed problem 2 reformulation N=1}.

Now, {let us show} that the limit point
$\{\Qb_{k}^{\star},t_{k}^{\star},\lambda_{k}^{\star}\}_{k=1}^{K}$ is
Pareto optimal to problem \eqref{designed problem 2 reformulation
N=1} {by contradiction}. Suppose that the limit point is not Pareto
optimal to problem \eqref{designed problem 2 reformulation N=1}.
Then, according to the definition {of Pareto optimality} in
\cite{Rubinov2001}, there exists a feasible solution to problem
\eqref{designed problem 2 reformulation N=1} such that $U({\tilde
t}_{1},\ldots,{\tilde
t}_{K})>U(t_{1}^{\star},\ldots,t_{K}^{\star})$, where ${\tilde
t}_{k}> t_{k}^{\star}$ and ${\tilde t}_{\ell}\geq t_{\ell}^{\star}$
for $\ell\neq k$. That is to say, the following constraint set is
feasible [by Lemma \ref{Lemma: eigenvalue decreasing}]:
\begin{subequations}\label{Proof: feasible set}
\begin{align}
&~\Phib_{k}\big(t_{k}^{\star}, \lambda_{k},
\{\Qb_{i}\}_{i=1}^{K}\big)
\succ \zerob, \label{Proof: feasible set 1}\\
&~\Phib_{\ell}\big(t_{\ell}^{\star}, \lambda_{\ell},
\{\Qb_{i}\}_{i=1}^{K}\big) \succeq \zerob,
~\forall \ell\neq k, \label{Proof: feasible set 2}\\
&~\sum_{k=1}^{K}\tr(\Qb_{k})\leq P, \label{Proof: feasible set 3}\\
&~\lambda_{k}\geq 0, ~\Qb_{k}\succeq\zerob, ~k=1,\ldots,K.
\label{Proof: feasible set 4}
\end{align}
\end{subequations}
{The feasibility of constraint \eqref{Proof: feasible set} implies
that there exists a solution $\{\lambda_{i},\Qb_{i}\}_{i=1}^{K}$
such that the minimum eigenvalue of the matrix
$\Phib_{k}(t_{k}^{\star}, \lambda_{k}, \{\Qb_{i}\}_{i=1}^{K})$ is
positive, rendering $z_{k}^{\star}>0$ in problem \eqref{decomposed
problem 1}, which contradicts with the fact of $z_{k}^{\star}=0$ by
Lemma \ref{Lemma: eigenvalue vs t}. Therefore, we have completed the
proof that the limit point yielded by Algorithm \ref{Algo: proposed
algorithm} is Pareto optimal to problem \eqref{designed problem 2
reformulation 2}.} \hfill $\blacksquare$

\section{Simulation Results}\label{Sec: simulation}

We consider the wireless system as described in Section~\ref{Sec:
system model} {with $N_{t}=4$ transmit antennas at the BS and $K=2$
single-antenna users, the total transmit power $P=10$ dB, and the
users' noise powers $\sigma_{1}^{2}=\sigma_{2}^{2}=0.01$.} For
simplicity, the CSI error radii of all users are assumed to be
identical, i.e., {$r_{1}=\cdots=r_{K}\triangleq r$}. In each
simulation trial, the presumed channels $\{{\hat
\hb}_{k}\}_{k=1}^{K}$ are randomly and independently generated
according to the standard complex Gaussian distribution.

\begin{figure}[t]
\vspace{-0.4cm}
\begin{center}
    \resizebox{0.53\textwidth}{!}{
    \psfrag{r}[Bc][Bc]{\large $r$}\hspace{-0.8cm}
        \includegraphics{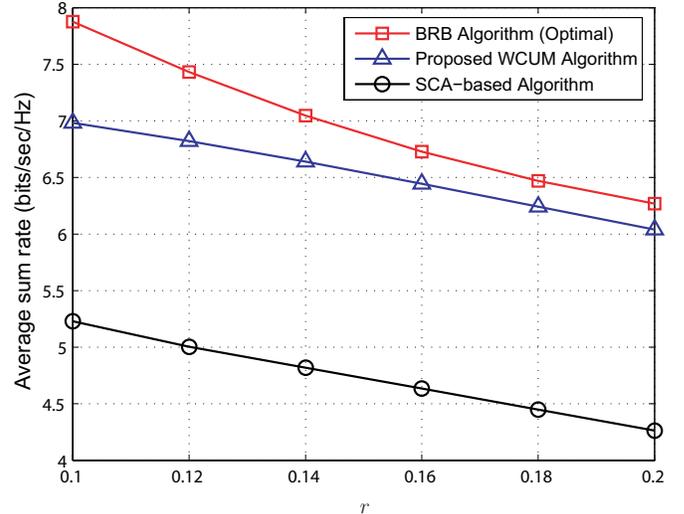}} 
        \caption{Performance comparison of the proposed WCUM algorithm and
        two existing algorithms in terms of average sum rate versus CSI
        error radius $r$.}\label{Worst_Rate}\vspace{-0.5cm}
\end{center}
\end{figure}

Considering sum-rate utility for problem \eqref{designed problem 2
reformulation 2}, we compare {the worst-case sum rate performances
of the proposed WCUM algorithm (Algorithm \ref{Algo: proposed
algorithm}), SCA-based algorithm \cite{Tran2012} and the optimal
(i.e., maximum) worst-case sum rate obtained by BRB algorithm (which
is a brute force approach) reported in \cite{Bjornson2012}.} The
solution accuracy for Algorithm \ref{Algo: proposed algorithm} and
SCA-based algorithm is set to $10^{-3}$, i.e., $\epsilon=10^{-3}$,
and the gap tolerance between the upper and lower bounds for the BRB
algorithm is set to 0.1 as in \cite{Bjornson2012}. The {involved}
convex problems in the algorithms {under test} are solved using
\texttt{CVX} \cite{cvx}. Figure \ref{Worst_Rate} shows the average
sum rate versus CSI error radius $r$, where the sum rates obtained
from the {three} algorithms are averaged over 100 {realizations} of
the presumed channels $\{{\hat \hb}_{k}\}_{k=1}^{K}$. {From this
figure, one can see that the proposed algorithm performs much better
than SCA-based algorithm, with the performance gap about 1.8
bits/sec/Hz, but worse than BRB algorithm, with the performance gap
between 0.8 bits/sec/Hz for $r=0.1$ and 0.2 bits/sec/Hz for $r=0.2$.
Note that the sum rate yielded by the proposed algorithm is closer
to the optimal (i.e., maximum) sum rate as the error bound $r$
grows, and that the performance gap between SCA-based algorithm and
BRB algorithm is as high as 2.6 bits/sec/Hz for $r=0.1$ and 2
bits/sec/Hz for $r=0.2$.}

{By our simulation experiences, we found that the proposed algorithm
converges much faster than BRB algorithm and is more computationally
efficient than SCA-based algorithm. As an illustration, the average
computation times of BRB algorithm, SCA-based algorithm, and the
proposed algorithm for obtaining the results shown in Fig.
\ref{Worst_Rate} for $r=0.1$ using a desktop PC with 3GHz CPU and
8GB RAM, are 550.9 secs, 123.9 secs, and 29.6 secs, respectively.
The fast convergence rate of the proposed algorithm is due to the
use of simple bisection method for solving problem \eqref{decomposed
problem 2}, and due to the smaller size of problem \eqref{decomposed
problem 1} compared with those problems involved in BRB and
SCA-based algorithms. {Let us emphasize that for large values of
$N_{t}$ and $K$, BRB algorithm will not be applicable due to
extraordinarily high complexity.} However, the detailed computation
complexity analysis {of Algorithm \ref{Algo: proposed algorithm}} is
omitted here due to the space limitation.}

\section{Conclusion}

We have presented a low computation complexity algorithm [see
Algorithm \ref{Algo: proposed algorithm}] for finding a {more
accurate} suboptimal solution for the {WCUM} problem in
\eqref{designed problem 2}. The proposed algorithm {has been proved}
to converge, and the limit point is Pareto optimal to the problem
[see Proposition~\ref{Prop: Pareto optimal}]. {The presented
simulation results have demonstrated that the proposed algorithm
performs much better than SCA-based algorithm and has higher
computational efficiency over both BRB and SCA-based algorithms.}

\section*{Acknowledgments}

This material is based upon works supported by the National Science
Council, R.O.C. under Grant NSC-99-2221-E-007-052-MY3, and by the
National Science Foundation under Grants 1147930 and 0917251.

\bibliography{References}

\end{document}